\documentclass{aa}  

\usepackage{natbib}
\bibpunct{(}{)}{;}{a}{}{,} 
\usepackage{graphicx}
\usepackage{subfigure}
\graphicspath{ {images/} }
\usepackage{txfonts}

\begin{document} 

\title{Identifying quasars with astrometric and mid-infrared methods from APOP and ALLWISE\thanks{Table 2 is only available in electronic form at the CDS via
anonymous ftp to cdsarc.u-strasbg.fr (130.79.128.5) or via
http://cdsarc.u-strasbg.fr/viz-bin/qcat?J/A+A/}} 
 \author{Sufen~Guo \inst{1,2} \and Zhaoxiang~Qi\inst{1,2}\thanks{\email{kevin@shao.ac.cn}} \and Shilong~Liao\inst{1,2} \and Zihuang~Cao\inst{2,4}\and Mario~G.~Lattanzi\inst{3} \and Beatrice~Bucciarelli\inst{3} \and Zhenghong~Tang\inst{1,2} \and Qing-Zeng Yan\inst{1,2}}

   \institute{Shanghai Astronomical Observatory, Chinese Academy of Sciences, 80 Nandan Rd, 200030 Shanghai, China
         \and School of Astronomy and Space Science, University of Chinese Academy of Sciences,  Beijing 100049, China
         \and INAF, Astrophysical Observatory of Torino, Via Osservatorio 20, 10025 Pino Torinese (Torino), Italy
         \and National Astronomical Observatory, Chinese Academy of Sciences\\ }

   \date{Received March 30, 2018; accepted June 10, 2018}
\titlerunning{Identifying quasars}
\authorrunning{Sufen et al.}
 
  \abstract
   {Quasars are spatially stationary, and they are essential objects in astrometry when defining reference frames. However, the census of quasars is far from complete. Mid-infared colors can be used to find quasar candidates because AGNs show a peculiar appearance in mid-infrared color, but these methods are incapable of separating quasars from AGNs.}
   {The aim of our study is to use astrometric and mid-infrared methods to select quasars and get a reliable quasar candidates catalog.}
   {We used a near-zero proper motion criterion in conjuction with WISE (all-sky Wide-field Infrared Survey Explorer) $[W1-W2]$ color to select quasar candidates. The $[W1-W2]$ color criterion is defined by the linear boundary of two samples: LAMOST DR5 quasars, which serve as the quasar sample, and LAMOST DR5 stars and galaxies, which serve as the non-quasar sample. The contamination and completeness are evaluated.}
   {We present a catalog of 662,753 quasar candidates, with a completeness of about 75\% and a reliability of 77.2\%.}
   {}


   \maketitle
%

\section{Introduction}

Quasars, one type of active galactic nuclei (AGNs), are point-like and generally at high redshifts. They are vital objects to construct the International Celestial Reference Frame (ICRF), which is the realization of the International Celestial Reference System (ICRS). This is based on the fact that the proper motions of quasars caused by secular abberration drift \citep{titov2010secular} is non-detectable now, and therefore they are regared as zero, which can keep the reference direction fixed. Consequently, quasars are ideal objects with which to construct a non-rotating reference frame. The International Astronomical Union (IAU) stipulated the precise coordinates of extragalactic radio sources as the ICRF \citep{arias1995extragalactic}. The first (ICRF1, \citealp{ma1998international}) and the second realization (ICRF2, \citealp{fey2015second}) of ICRF are defined by the positions of compact radio sources (mostly quasars) obtained with VLBI. With the improvement of precisions and the spatial and frequency coverages at radio band, ICRF3 \citep{christopher2014icrf,malkin2015icrf} will be released in 2018, containing a larger number of radio-loud quasars.  

Radio-loud quasars that have optical counterparts can be used to tie ICRF to optical frames \citep{bourda2011vlbi}. However, this alignment requires extensive overlapping sources, which possess highly accurate radio as well as optical positions and a uniform sky coverage. European Space Agency's (ESA) Gaia mission \citep{lindegren2016gaia,mignard2016gaia} is realizing a new optical reference frame. This mission is dedicated to measure three-dimensional positions and velocities with unprecedented precision for more than 1 billion objects within five years. Observations will contain $\sim$500,000 quasars, 20,000 of which are brighter than visual magnitude 18. The first (Gaia DR1, \citealp{lindegren2016gaia}) and second (Gaia DR2, \citealp{2018yCat.1345....0G}) datasets have been released, and the final data release is expected to be published by 2020. Unlike the preceding Hipparcos mission \citep{perryman1997hipparcos}, Gaia realizes the extragalactic celestial reference frame directly at optical wavelengths, using the highest accurate positions of quasars \citep{lindegren2016gaia,mignard2016gaia,2018arXiv180409366L}. It is crucial to guarantee the consistency between this optical reference frame and the radio reference frame. To achieve this goal, a large number of quasars are needed to produce a uniformly distributed sample and to place severe constraints on the overall geometry.

Many surveys have been performed to complete the census of quasars, and the number of quasars now is approximately 700,000. The main surveys include the Sloan Digital Sky Survey (SDSS, \citealp{york2000sloan}), the catalog from the large Sky Area Multi-Object Fiber Spectroscopic Telescope (hereafter LAMOST, \citealp{1674-4527-12-9-003}), the Large Bright Quasar Survey \citep{hewett1995large}, the Hamburg Quasar Survey \citep{hagen1995hamburg}, the Canada-France High-z Quasar Survey \citep{1538-3881-139-3-906}, the INT Wide Angle Survey \citep{sharp2001first}, and the 2dF Quasar Survey \citep{boyle20002df}. The SDSS data release series \citep{paris2012sloan,ross2012sdss,paris2014sloan,myers2015sdss,parissloan2017b,parissloan2017a} contribute much of the effort.

Although a large number of quasars have been confirmed by their spectra, many quasars have not yet been identified. Recently, several selection methods for AGN candidates using photometric observations have been developed, and among them, the mid-infrared color criteria, which are based on the fact that AGNs occupy a specific locus in the mid-infrared color space \citep{lacy2004obscured,richards2006spectral}, are proved to be very efficient. These methods can efficiently select AGNs, but not quasars. 

Mid-infrared selection methods are relatively insensitive to extinction caused by interstellar dust which plagues AGN surveys toward shorter wavelengths. Notably, with the release of the mid-infrared photometric data of all-sky Wide-field Infrared Survey Explorer (WISE, \citealp{wright2010wide}), some selection criteria using WISE colors have been developed. For instance, \citet{stern2012mid} using $[W1-W2]$ color identified 78\% AGNs from AGN candidates with 95\% reliability; \citet{mateos2012using} defined a criterion with $[W1-W2]$ and [W2-W3] colors. However, the reason why AGNs are located in particular locus in mid-infrared color space is not yet fully understood. \citet{stern2005mid,stern2012mid} assumed that in the mid-infrared band, the spectral energy distribution (SED) of AGNs are dominated by a power law, which is different from blackbody spectrum objects like stars and normal galaxies. Alternatively, \citet{nikutta2014meaning} claims that the difference in mid-infrared results from the peculiar characteristics of the dust shells surrounding different type of objects.

With mid-infrared methods, \cite{secrest2015identification} selected 1.4 million AGN candidates (hereafter, MIRAGN).This large sample of AGNs together with a tiny fraction of VLBI sources was used to define the reference frame of Gaia DR2 \citep{2018arXiv180409377M}. Due to the deeper WISE coverage, the objects' density in MIRAGN near the ecliptic poles is much larger than that in other regions (see detail in Fig.~1 of \citealp{secrest2015identification}). Furthermore, in MIRAGN, about 70,000 candidates have proper motions larger than 10 ${\rm mas\; yr^{-1}}$ either in right ascension or in declination, suggesting that they are probably stars or extended galaxies. This may lead to the instability of Gaia reference frame, and it is therefore necessary to test this reference frame with a uniformly distributed quasar sample.

Non-detectable proper motion is a basic characteristic of quasars and a constraint on this can purify quasar candidates selected in mid-infrared color maps. In fact, because most proper motion catalogs are derived from optical observations, the astrometric criterion can eliminate most dwarfs that are ultracool but have high mid-infrared colors (see Fig.~1 in \citealp{kirkpatrick2011first}). However, the proper motion has not been widely used as a selection criterion because of lacking  precise proper motions. The catalog of the Absolute Proper motions Outside the galactic Plane (APOP, \citealp{qi2015absolute}), which contains precise proper motions, makes it feasible to identify quasars in such a way. APOP covers $\sim$12,000 square degrees of the southern sky and the limiting magnitude $R_{F}$ is 20.8, so it includes many dark sources beyond the regime of SDSS.

In order to derive a reliable sample of quasars, we present a quasar selection method by combining mid-infrared criterion with a near-zero proper motion criterion. This method is the first attempt to use both astrometric and mid-infrared data, and can be applied to subsequent catalogs that contain precise proper motions. 

This paper is organized as follows. We describe the data used in Section \ref{data} and the detailed quasar selection criteria in Section \ref{selection}. In Section \ref{results} we present our quasar candidate catalog. We describe the catalog in Secion \ref{catalog}. Finally, in Section \ref{conclusion} we summarize our results.
  
\section{Data}\label{data}
\subsection{The APOP catalog}\label{data:apop} 
APOP is derived from historical digitized Schmidt survey plates, which covers 22,525 square degrees of the sky where $|b|\geq 27$ degrees. Using stars and galaxies, APOP removed the systematic errors in proper motions related to positions, magnitudes and colors. Additional $R_{F}$, $B_{J}$, $I_{V}$ and $V$ magnitudes extracted from GSC 2.3 \citep{1538-3881-136-2-735} and $JHK$ magnitudes from 2MASS \citep{skrutskie2006two} are listed. The reference frame is aligned with the ICRF2 within $\pm 0.2$ mas at J2000.0. More than a hundred million sources are cataloged and the magnitude reaches $R_{F}\sim 20.8$. The internal accuracy of proper motion is better than 4.0 ${\rm mas\; yr^{-1}}$ for stellar objects brighter than  $R_{F}$= 18.5, and 9.0  ${\rm mas\; yr^{-1}}$ for stellar objects with magnitudes $18.5<R_{F}<20.0$. The nominal global zero point error estimated by quasars is better than 0.6 ${\rm mas\; yr^{-1}}$.

\subsection{The ALLWISE catalog}\label{data:allwise}
WISE is a satellite with a 40 cm aperture launched by NASA in 2009 December. It scanned the whole sky at four mid-infrared bands, centering at 3.4, 4.6, 12 and 22 $\mu$m (hereafter referred to as W1, W2, W3, and W4, respectively) with image resolutions (FWHM) of 6.1, 6.4, 6.5, and 12.0 arcsec. The achieved sensitivities in the preliminary data release are about 0.008, 0.11, 1 and 6 mJy ($5\sigma$) for the four bands, respectively. Began in 2010 January, WISE finished the first all-sky cryogenic survey in August 2010. After that, an additional $30\%$ of the sky was mapped. The final post-cryogenic survey (NEOWISE, \citealp{mainzer2011preliminary, 0004-637X-731-1-53}) started on September 2010 and finished on February 2011, covering $70\%$ of the sky. 

The ALLWISE catalog \citep{2013yCat.2328....0C} is a compilation of the data from the cryogenic and post-cryogenic survey phases of the WISE mission. It contains more than seven hundred million celestial objects with improved photometric and astrometric accuracy compared with the WISE All-Sky Release Catalog \citep{2012yCat.2311....0C}. ALLWISE catalog lists positions, magnitudes, astrometric and photometric uncertainties, flags, and counterparts in the 2MASS catalog.  

\subsection{The LAMOST catalog}\label{data:lamost}
LAMOST is a reflecting Schmidt telescope located at Xinglong Observing station in China. With a wide field of view (FOV) of $5^\circ$ that can allocate 4000 optical fibers, it can obtain 4000 spectra simultaneously. A large aperture with diameter of about 4 m is designed to observe deep objects. It is dedicated to observe the spectrum of celestial objects over the northern sky with resolution $R=\lambda / \bigtriangleup \lambda=$ 1800. The limiting magnitude in the $i$ band is 20.5. LAMOST began its first spectroscopic survey in September 2012, and it has released the fifth data release (LAMOST DR 5) with more than 1.3 million spectra, containing 1.2 million stars, 36093 galaxies and 14,782 quasars. Before DR5, LAMOST used a method similar to SDSS to identify quasars \citep{1674-4527-15-8-1095}, while for DR5, quasars were confirmed by human eyes, making identifications more reliable.

\section{Quasar selection}\label{selection}
We used mid-infrared colors in conjunction with a near-zero proper motion criterion to select quasars. In this section, we discuss in detail our selection criteria of quasar candidates.

\subsection{Mid-infrared color-selection criterion}\label{selection:color}
With the rise of mid-infrared surveys, researchers discovered that  AGNs primarily reside in a restricted locus in mid-infrared color space, which allows them to separate AGNs from stars and normal galaxies in mid-infrared color maps. For the first time, \cite{lacy2004obscured} and \cite{stern2005mid} investigated AGNs with mid-infrared color using the Spitzer Space Telescope First Look Survey, and since then, the development of mid-infrared method has profited significantly from the successful WISE mission. After the WISE mission was completed, researchers \citep{stern2012mid,secrest2015identification,assef2011mid} found that WISE colors are efficient to select AGNs. Mainly there are two methods that rely {\em only} on colors from WISE imaging: one uses the $[W1-W2]$ color only \citep{stern2012mid}; the other was developed by \citet{mateos2012using}: using two WISE colors ( $[W1-W2]$ and $[W2-W3]$ ) to define the boundaries of the region where AGNs are located. All these criteria generally agree with each other. The majority of the spectroscopically confirmed or X-ray confirmed AGNs were used to test the reliability of these methods. 

\begin{figure}[htp]
\includegraphics[width=0.48\textwidth]{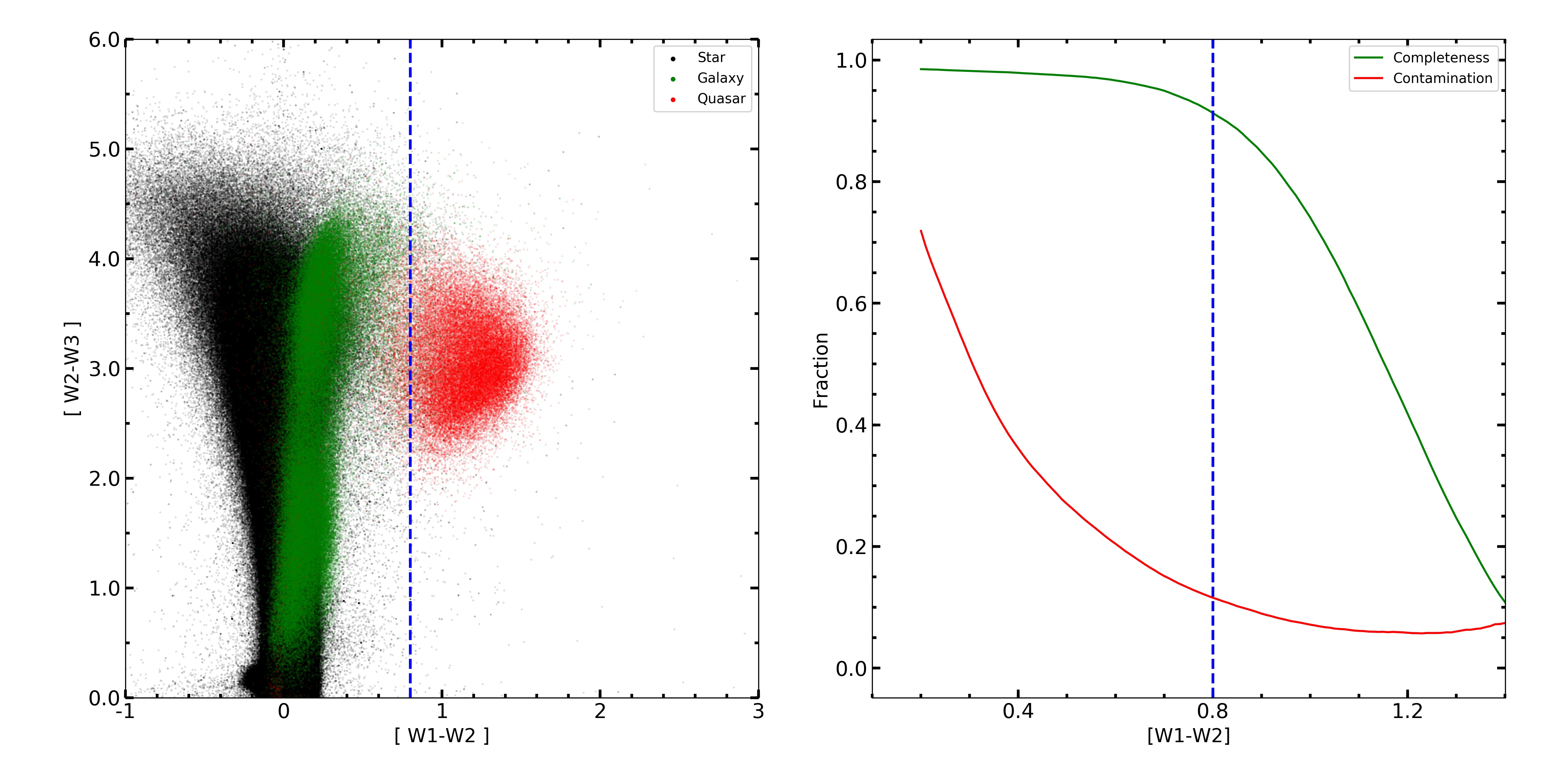}
\caption{Left panel: Quasar locus in [W1-W2] and [W2-W3] space. Right panel: Quasar completeness (blue curve) and the contamination (red curve) by stars and galaxies as a function of a $[W1-W2]$ color cut. The dotted lines in the two panels cut at $[W1-W2]=0.8$.}
\label{cc}
\end{figure}

\begin{figure}[htp]
\includegraphics[width=0.48\textwidth]{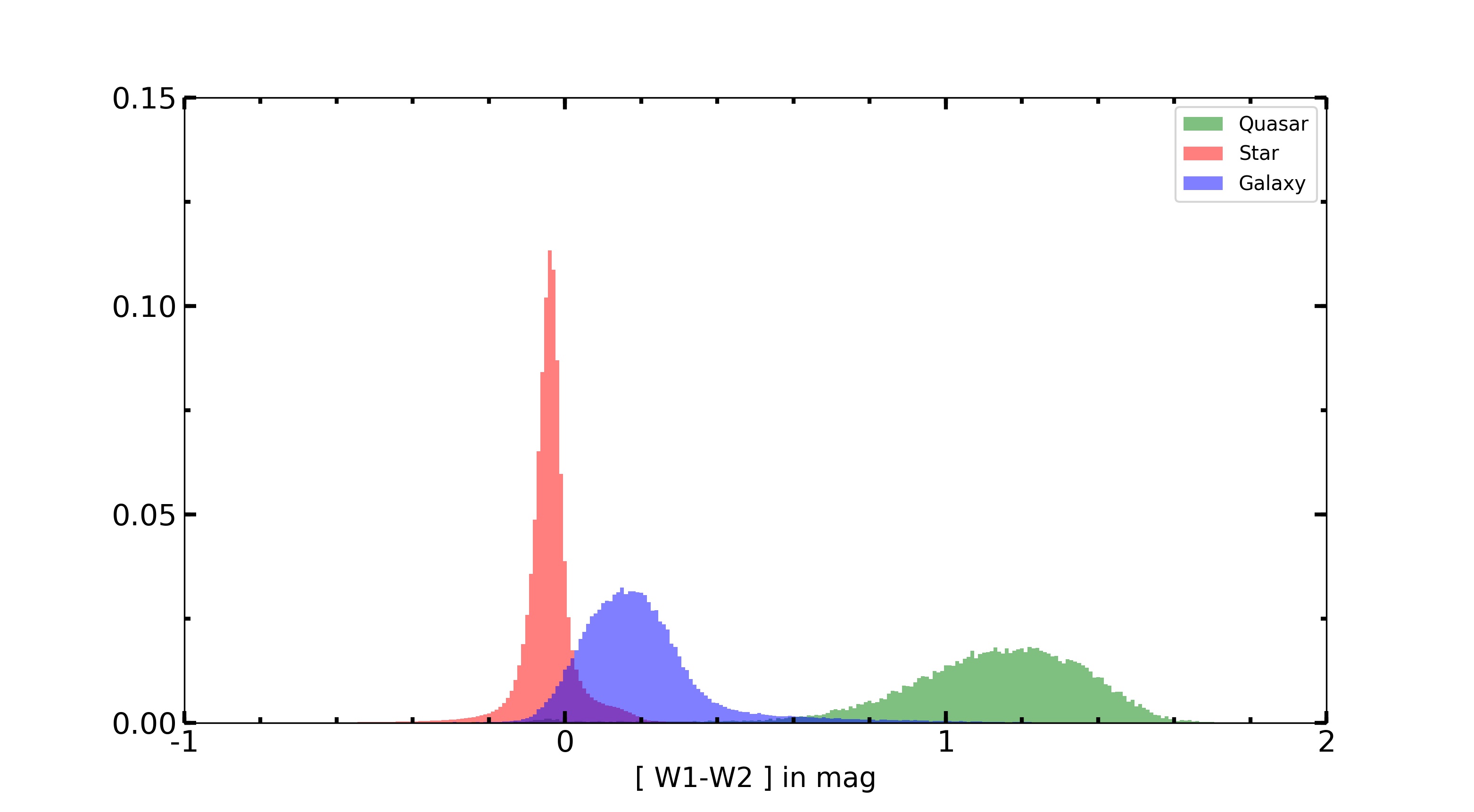}
\caption{Normalized histogram of $[W1-W2]$ color of quasars (green), stars (red) and galaxies (blue).}
\label{hist1}
\end{figure}

Building upon these valuable works, we carried out an independent investigation of mid-infrared selection criteria using new data. Because the newly released LAMOST DR5 has a highly reliable sample of quasars, we used it to define our selection criteria. 8,169,852 common sources are found in ALLWISE and LAMOST DR5, including quasars, galaxies and stars. We used SVMs (support vector machines) to determine the color boundaries. The SVM algorithm is a widely used technique capable of classifying objects into different groups in a supervized machine-learning environment; in particular, linear SVM classifiers are suitable in this context, as previous works \citep{stern2012mid,mateos2012using} showed that AGNs are linearly separable from other objects. The SVM method is performed with the $Python$ package $sklearn$, in which the penalty value $C$ is a tunable parameter: a larger $C$ means higher penalties for misclassification whereas a smaller $C$ allows more misclassification. Because we want our quasar sample to be reliable, we chose an intermediate value of $C$ =10. We investigated two-color selection criteria, i.e., $[W1-W2]$ and $[W2-W3]$, and single-color selection criteria, i.e., $[W1-W2]$. We found that single-color criterion work better on classification. As illustrated in the left panel of Fig.~\ref{cc}, quasars can be efficiently isolated from stars and galaxies by a simple $[W1-W2]$ cut. The right panel of Fig.~\ref{cc} shows that the contamination declines at the cost of losing completeness, and $[W1-W2]\geq 0.8$ is a balance between low contamination and high completeness. The histogram of quasars, stars and galaxies in Fig.~\ref{hist1} clearly shows this separation. Accordingly, we defined the mid-infrared criterion as: $[W1-W2]\geq 0.8$, in agreement with \cite{stern2012mid}. With this criterion, we find that the completeness of LAMOST quasars is $91.4\%$ and the contamination by sources that are not quasars is $11.1\%$. Additionally, in order to obtain a reliable sample of mid-infrared photometry, we rejected sources from ALLWISE having signal-to-noise ratios (S/N) lower than five in the W1 or W2 bands. 

\begin{figure*}[htp]
\includegraphics[width=1\textwidth]{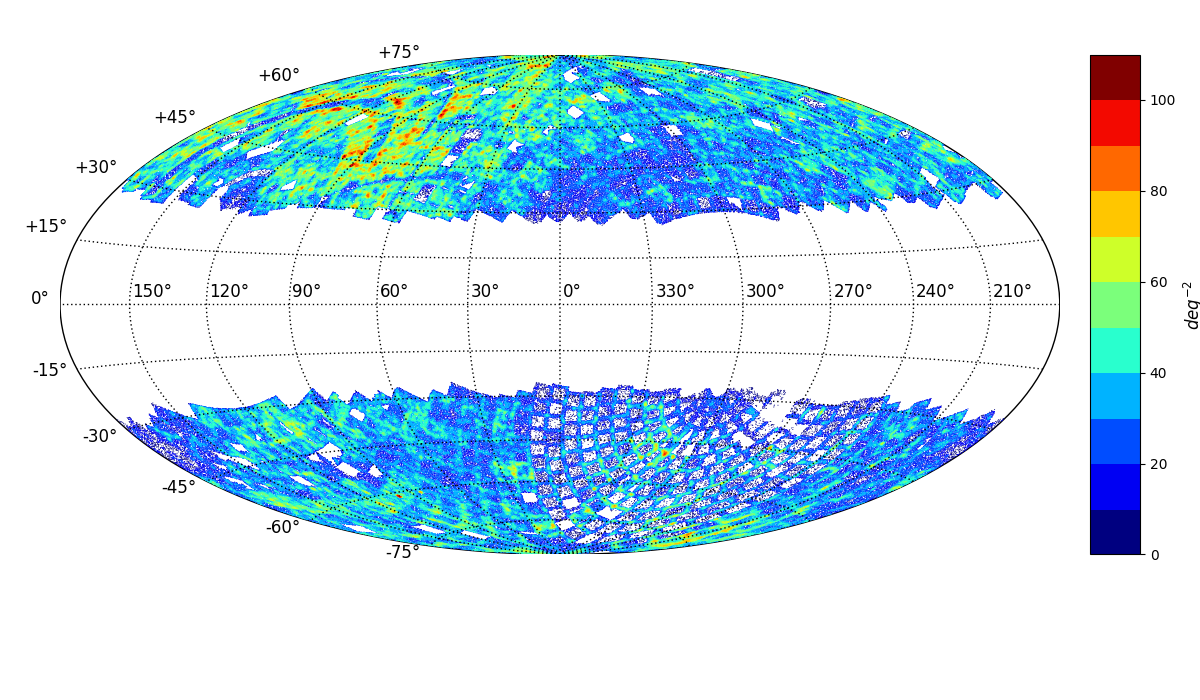}
\caption{Sky density plot for the selected quasar candidates of this work, Aitoff projection in galactic coordinates. The sources are outside the Galactic plane by virtue of the quasar proper motion criteria which used APOP catalog.}
\label{distribute}
\end{figure*}

\subsection{Proper motion criteria}\label{selection:pm}
Due to their distant extragalactic distances, the proper motion of quasars can be considered to be zero. On the other hand, ground-based proper motion catalogs suffer from low-precision data and/or systematic error sources and, consequently, are affected by zero-point error. Therefore, if we use the APOP proper motions to bulid a criterion for quasar selection, we need to consider the zero-point error of APOP first, and then we can define a suitable proper moption criterion.

The APOP catalog estimated the global zero-point using the quasars confirmed by the Gaia Initial QSO Catalog (GIQC), which contains about 1.2 million quasar candidates from LQAC2 \citep{souchay2012second}, SDSS DR7 \citep{1538-3881-139-6-2360}, 2dF, BOSS, VLBI observation and data from \citet{1538-3881-137-4-3884}. GIQC potentially included some spurious sources, resulting in an inaccurate zero-point. 

To calculate a global zero-point purely based on confirmed quasars, we cross-matched APOP with LAMOST DR5, obtaining 42,140 quasars. We rejected the sources with proper motion errors larger than 10 ${\rm mas\; yr^{-1}}$ either in right ascension or declination. With $2\%$ outlier rejection, we obtained mean proper motions of $0.0\pm 4.6$ and $0.5\pm 4.4$ ${\rm mas\; yr^{-1}}$ in right ascension and declination, respectively.

The newly released third Large Quasars Astrometric Catalog (hereafter LQAC3, \citealp{souchay2015third}) includes most of the confirmed quasars. We cross-matched APOP with LQAC3 and estimated the zero-point as a comparison. There are 185,197 sources in common. With the same procedure, we got mean proper motions of $0.3\pm 4.4$ and $0.6\pm 4.0$ mas $\rm yr^{-1}$ in right ascension and declination, respectively.

We adopted the zero-point error derived from LAMOST DR5. Although LQAC3 shows a similar result, quasars in LAMSOT DR5 are more reliable, because they have been confirmed by visual inspection. Finally, in order to take into account the astrometric signature of quasars, we removed objects whose proper motions in right ascension or declination deviate from the mean APOP zero-point error by more than $3\sigma$; this is not a strict criterion but one deemed necessary to exclude most of the contamination of stars inside the Milky Way and extended AGNs. Furthermore, sources with APOP proper motion errors in right ascension or declination larger than 10 ${\rm mas\; yr^{-1}}$ were excluded.

\section{Results}\label{results}
\begin{figure}[htp]
\includegraphics[width=0.5\textwidth]{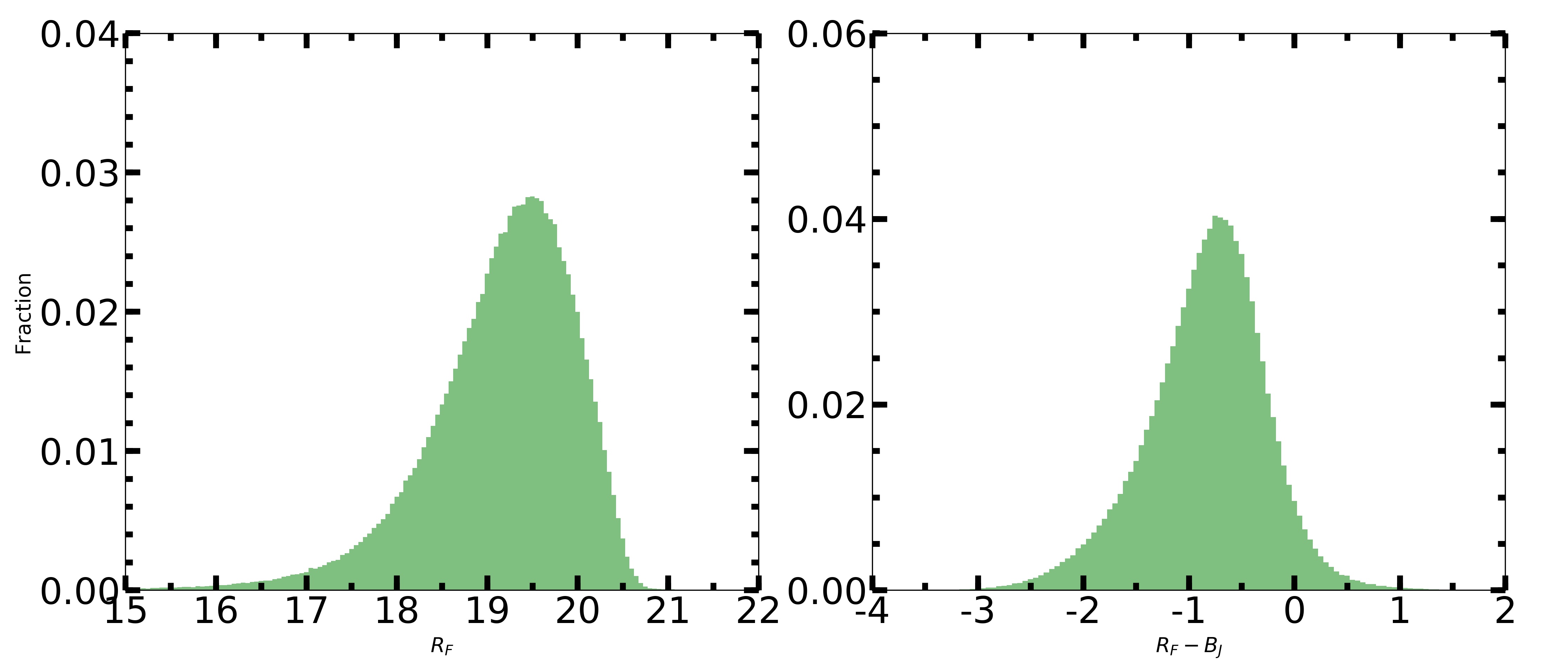}
\protect\caption[position=bottom]{Normalized histograms of $\rm R_{F}$ magnitudes and $\rm R_{F}-B_{J}$ color.
The left panel is the histogram of the $\rm R_{F}$ magnitudes of sources in our catalog, which shows a peak at about $\rm R_{F}$=19.5. The right panel shows the $\rm R_{F}-B_{J}$ color.}
\label{hist_mag}
\end{figure}

We obtain a catalog of 662,753 quasar candidates outside the Galactic plane, and 488,130 are new quasars that were not identified either by LAMOST, SDSS, or X-ray observations. The mean source density is about 30 $\rm deg^{-2}$, covering 22,525 square degrees. Figure~\ref{distribute} shows the density plot of our candidates. Except for the region near the ecliptic poles, it is relatively uniformly distributed outside the Galactic plane. The reason causing the non-uniformity near the ecliptic poles is the deeper WISE coverage toward the vicinity of this region. This problem is also present in MIRAGN, but our results have been improved because we imposed extra restrictions on proper motions. In Table~\ref{table:col} we detail the content of our catalog. The histogram of $\rm R_{F}$ magnitudes is shown in the left panel of Fig.~\ref{hist_mag} and the right panel of Fig.~\ref{hist_mag} is the histogram of $\rm R_{F}-B_{J}$ color. We excluded the candidates that lack $\rm R_{F}$ or $\rm B_{J}$ magnitude values, and in total, about 66,000 candidates remain in this plot.

\tabcolsep=0.11cm
\begin{table}      
\centering               
\begin{tabular}{c c c}      
 \hline\hline                 
Resource & Completeness & Quasars found/All quasars \\
\hline           
   ICRF2        &84.6\% &1057/1250 \\ 
   VLBA         &70.8\% &1731/2446 \\ 
   VLA          &87.0\% &613/705\\
   JVAS         &73.3\% &631/861 \\
   FIRST        &92.9\% &769/828\\
   2QZ          &79.5\% &12279/15439\\
   SDSS DR7-DR10 &81.0\% &96014/118521\\
   2df-SDSS LRG &69.5\% &1477/2125\\
   GSC2.3       &84.0\% &86312/102806\\
   B1.0         &84.2\% &86280/102519\\
   2MASS        &77.3\% &14071/18207\\
   HB           &84.9\% &4043/4762\\
   V\&V         &32.2\% &3765/11701\\
\hline                       
\end{tabular}
\caption{Completeness compared with LQAC3}
\label{table:1}
\end{table}
\begin{figure} 
\centering 
\subfigure[]{
  \label{fig:subfig:a}
  \includegraphics[width=0.45\linewidth]{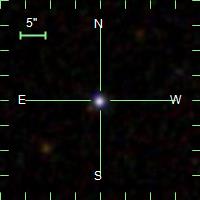}
  }
\hspace{0.01\linewidth}
\subfigure[]{
  \label{fig:subfig:b}
  \includegraphics[width=0.45\linewidth]{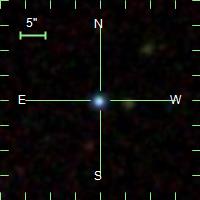}
  }
\vfill
\subfigure[]{
  \label{fig:subfig:a}
  \includegraphics[width=0.45\linewidth]{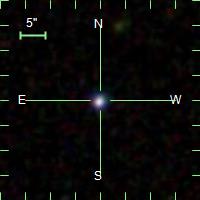}
  }
\hspace{0.01\linewidth}
\subfigure[]{
  \label{fig:subfig:b}
  \includegraphics[width=0.45\linewidth]{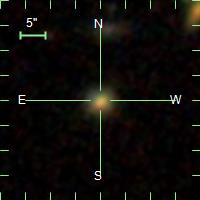}
  }
\caption{Four sources taken from our catalog matching with SDSS DR14: (a), (b), and (c) show point-like sources and (d) shows an extended source.}
\label{fig:mor}
\end{figure}

\subsection{Completeness}\label{results:completeness}
We used the compiled quasar catalog LQAC3 \citep{souchay2015third} to estimate the completeness of our sample. In total, 185,197 common sources are found in all three catalogs: LQAC3, APOP and ALLWISE. Of these, 148,138 sources have both clear ALLWISE detections (S/N in W1 and W2 > 5) and well-qualified proper motions (error < 10 ${\rm mas\; yr^{-1}}$). With this robust sample, we calculated the completeness of the sources going back to the raw sources of LQAC3 to check how many quasars are found. The result is shown in Table~\ref{table:1}. Because we added a proper motion criterion to the mid-infrared method, as expected, the completeness is lower than the catalogs which use mid-infrared selection method purely. All the completeness is around 75\% but for the result from \citet{veron2010catalogue} (hereafter V\&V) catalog. Among the 11,701 common sources from APOP and V\&V, 7936 sources in V\&V catalog are absent in our final catalog. We found that the $[W1-W2]$ value of most of the sources (7648) in V\&V catalog is smaller than 0.8 (rejected by our mid-infrared selection criterion), the rest 288 sources have proper motion errors larger than 10 ${\rm mas\; yr^{-1}}$ (rejected by our proper motion criterion).

\subsection{Contamination}\label{results:contamination}
To evaluate the contamination by stars and galaxies, we randomly chose a $30^{\circ}\times 30^{\circ}$ testing region. Within this region, we compiled all sources with confirmed classification (star, galaxy, quasars) from SDSS DR14, LAMOST DR5 and LQAC3 and take it as a reference catalog. Such catalog lists 8043 sources in common with our quasar candidate catalog, of which 7778 are confirmed quasars (96.7\%), 69 are confirmed stars (0.9\%) and 196 are confirmed galaxies (2.4\%). This suggests that only $\sim$3\% of our sources are not AGN-like. We note, however, that a reasonable fraction of the confirmed "quasars" in our sample are low-luminosity sources at low redshift. Such sources may have a visible host galaxy, and so be extended in imaging, limiting their utility for astrometric calibration.

Our purpose is to identify as many quasar candidates as possible, which means the main contents in our final catalog should be point-like sources. We investigated the photometric nature of our catalog by matching it to the SDSS imaging database to evaluate the extended sources contamination. We randomly chose 1000 sources and inspected their morphology using the SDSS tools. 632 sources are found in the SDSS imaging database, of which 488 are point-like (77.2\%), 144 are extended (22.8\%). Fig.~\ref{fig:mor} shows the SDSS thumbnails (http://skyserver.sdss.org/dr14/en/tools/chart/navi.aspx) of four candidates in our catalog.

The contamination by other type of AGNs is difficult to remove because of the inevitable proper motions errors in APOP. It is possible to exclude many more suprious sources, with the astrometry techique improved in the future. 

\subsection{Redshifts}\label{results:redshift}
\begin{figure}[htp]
\includegraphics[width=0.5\textwidth]{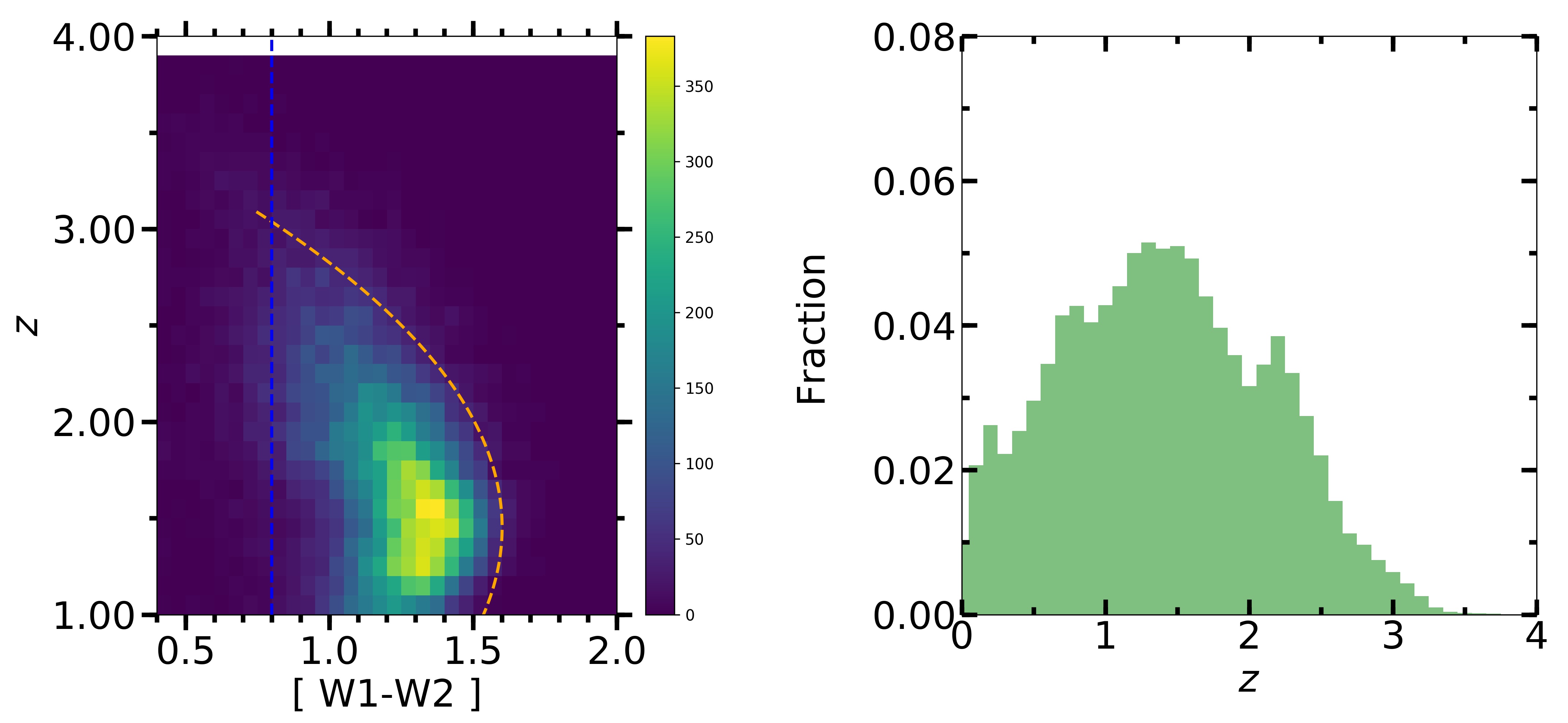}
\caption{Left panel: Redshifts of quasars against [W1-W2] color. 48,282 quasars with valid redshift from LAMOST DR5 are plotted. The dashed blue line is the [W1-W2] >=0.8 cut and dashed orange line is the fitting curve of the edge. Right panel: Normalized histogram of the redshift of quasar candidates in this work. Redshift value is extracted from SDSS DR14 (145,205 sources) and LAMOST DR5 (14,265 sources). In total, 159,470 valid redshifts of quasars are plotted.}
\label{plot_z}
\end{figure}
At high redshift ($z\,>\,3.5$), when the $\rm H\alpha$ line is redshifted to the W1 band or the $\sim$1 $\mu$m minimum enters the W2 band (see detail in \citealp{stern2012mid}), the $[W1-W2]$ color of AGNs may be bluer (i.e., smaller) than 0.8. The left panel of Fig.~\ref{plot_z} confirms this behavior. Therefore, our catalog loses some high-redshift quasars as all mid-infrared selection methods do. The right panel of Fig.~\ref{plot_z} is the histogram of the quasars which have valid redshifts from our catalog. Most of the redshifts are less than three and no significant high-redshift ($z\,>\,3.5$) quasars are identified. It should be noted that since the left panel of Fig.~\ref{plot_z} contains only the qusars from LAMOST, this figure may be biased by the observaiton strategy of LAMOST. 

\subsection{Comparison with MIRAGN}
\begin{figure}[htp]
\includegraphics[width=0.5\textwidth]{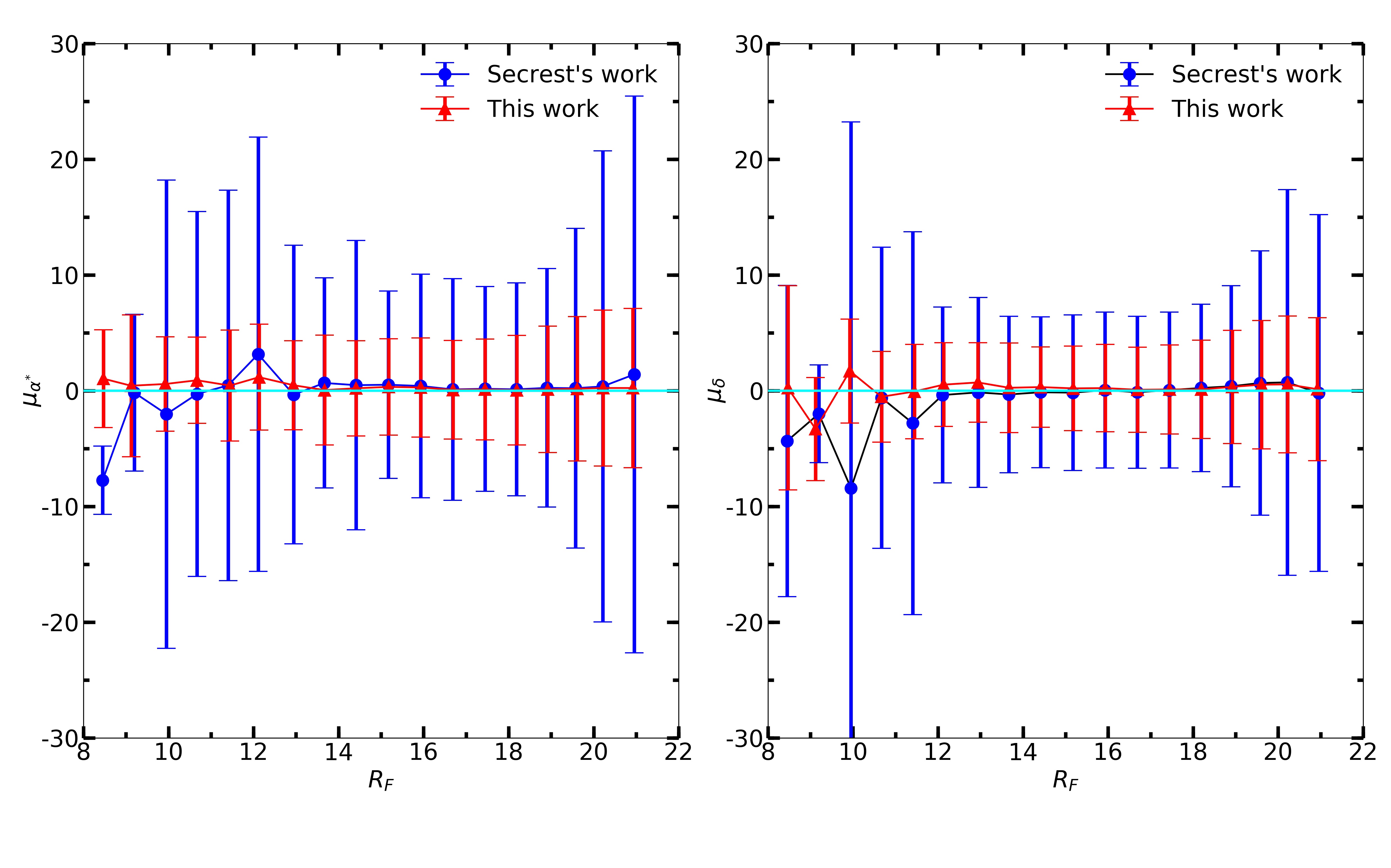}
\caption{Proper motions of quasars against $R_{F}$ magnitude. The red triangles indicate the mean proper motions of our catalog of $\alpha^{*}$ and $\delta$ in that magnitude bin and the error bar shows their standard deviations in ${\rm mas\; yr^{-1}}$. The blue circles indicate the mean proper motions of MIRAGNs in each 0.75-magnitude bin and the error bar indicates their standard deviations.}
\label{pmbin}
\end{figure}

Figure~\ref{pmbin} shows the proper motions of the MIRAGNs (\citealp{secrest2015identification}) and our quasar sample against $R_F$ magnitude and indicates that they are consistent with the zero-point error estimated by APOP. The red lines are the proper motions from our catalog and the blue lines from MIRAGN. The mean and dispersion of our work are closer to zero and shows less trend along $R_F$ magnitude. 

\section{Description of the catalog}\label{catalog}
The quasar candidate catalog presented in this paper contains 662,753 objects. The catalog is available at Strasbourg’s astronomical Data Center (http://cds.u-strasbg.fr, catalog ?); the description of the catalog data is detailed in Table~\ref{table:col} and we describe the items for each line of the catalog as follows. 
\begin{table*}
\centering                   
\begin{tabular}{c c c c c}      
 \hline\hline                 
Label&Type&Units&Detail  \\
\hline           
   ID1 & Int & - & ID1 of the unique union id (ID1,ID2)\\ 
   ID2 & Int & - & ID2 of the unique union id (ID1,ID2)\\
   RAdeg & Double & degree & Right Ascension in J2000.0\\
   DEdeg & Double & degree & Declination in J2000.0\\
   eRAmas & Double & mas & Error of right ascension\\
   eDECmas & Double & mas & Error of declination\\
   pmRA  & Double & ${\rm mas\; yr^{-1}}$ & $\mu_{\alpha}cos\delta$\\
   pmDEC & Double & ${\rm mas\; yr^{-1}}$ &  $\mu_{\delta}$\\
   epmRA & Double & ${\rm mas\; yr^{-1}}$ & Error of $\mu_{\alpha}cos\delta$\\
   epmDEC & Double & ${\rm mas\; yr^{-1}}$ & Error of $\mu_{\delta}$\\
   Rmag  & Float & mag & $R_F$ photographic magnitude\\
   Bmag  & Float & mag & $B_J$ photographic magnitude\\
   Vmag  & Float & mag & $V$ photographic magnitude\\
   Imag  & Float & mag & $I_N$ photographic magnitude\\
   Jmag  & Float & mag & 2MASS $J$ magnitude\\
   Hmag  & Float & mag & 2MASS $H$ magnitude\\
   Kmag  & Float & mag & 2MASS $K$ magnitude\\
   W1mag  & Float & mag & ALLWISE $W1$ magnitude\\
   W2mag  & Float & mag & ALLWISE $W2$ magnitude\\
   W3mag  & Float & mag & ALLWISE $W3$ magnitude\\
   W4mag  & Float & mag & ALLWISE $W4$ magnitude\\
   QSOflag  & Char & - & Flag on the presence of SDSS or LAMOST\\
   z  & Float & - & Redshift\\
   zflag  & Char & - & Flag on the sources of redshift\\
   Morflag  & Char & - & Flag on the morphology\\
\hline                       
\end{tabular}
\caption{Description of our catalog.}
\label{table:col}      
\end{table*}

\begin{itemize}
\item Columns 1 and 2 give the unique union identifier (ID1,ID2)
\item Columns 3 and 4 are the $\alpha$, $\delta$ coordinates given by the APOP catalog in epoch J2000.0 and their errors are in Column 5 and 6, respectively.
\item Columns 7 and 8 give the proper motions $\mu_{\alpha}cos\delta,\;\mu_{\delta}$ and in Columns 9 and 10 are their errors, respectively.
\item Columns 11, 12, 13, and 14 provide the $R_F$, $B_J$, $V$, $I_N$ magnitudes directly extracted from APOP.
\item Columns 15, 16, and 17 give the near-infrared J (1.25 $\mu$m), H (1.65 $\mu$m) and K (2.16 $\mu$m) magnitudes directly extracted from 2MASS catalog.
\item Columns 18, 19, 20, and 21 are the mid-infrared W1 (3.4 $\mu$m), W2 (4.6 $\mu$m), W3 (8 $\mu$m), W4 (22 $\mu$m) magnitudes directly extracted from ALLWISE.
\item Column 22 gives the flags indicating the presence of the quasar in SDSS (flag: "s") or LAMOST (flag : "g"). 
\item Column 23 gives the redshift value obtained from SDSS if QSOflag (Column 22) ="s"; from LAMOST if QSOflag (Column 22) = "g"; else z=-99 .
\item Column 24 gives the flags indicating the sources of redshift from SDSS (zflag="s") or from LAMOST (zflag = "g"); z="o" for no valid redshift.
\item Column 25 is the morphological flag indicating whether these objects are point-like or extend sources. "p"= point-like sources; "e" = extend sources.
\end{itemize}

\section{Conclusions}\label{conclusion}
We have presented a catalog of quasar candidates selected using mid-infrared and proper motion methods; 96.7\% of our sources are confirmed to be quasar-like, but not all of these quasars are point-like in imaging. We estimate that $\sim$22.8\% may be extended in imaging, limiting their utility for establishing an astrometric reference frame.

Newly released LAMOST data allowed us to investigate the $[W1-W2]\geq 0.8$ criterion to discriminate quasars from stars and galaxies, confirming previous studies based on SDSS data \citep{stern2012mid}.
Our sample lies outside the Galactic plane, allowing us to avoid many strongly polluted WISE images near the plane. Also, having used astrometric constraints, the contamination of brown dwarfs with detectable proper motions are also small. A tiny fraction of normal galaxies may be included in our catalog because of mid-infrared emission from star bursts or strong star formation regions in them.

Finally, although about a quarter of the sources in our catalog are extended in imaging, the near-zero proper motion criterion ensures the effectiveness of an external test of the reference frame of Gaia DR2 which was established mainly by MIRAGNs \citep{2018arXiv180409377M}. Our quasar candidates can be used to construct a uniformly distributed quasar catalog together with the confirmed quasars to analyze the parallax zero-point \citep{michalik2016quasars} of the Gaia DR2 and the link of the optical reference frame to the ICRS \citep{mignard2016gaia}.

\begin{acknowledgements}
We thank the referee for constructive scientific comments. 

This work is supported by grants from the National Science Foundation of China (NSFC No. 11573054, 11503042, 11703065). 

This publication makes use of data products from the Wide-field Infrared Survey Explorer, which is a joint project of the University of California, Los Angeles, and the Jet Propulsion Laboratory/California Institute of Technology, funded by the National Aeronautics and Space Administration. 

Guoshoujing Telescope (the Large Sky Area Multi-Object Fiber Spectroscopic Telescope LAMOST) is a National Major Scientific Project built by the Chinese Academy of Sciences. Funding for the project has been provided by the National Development and Reform Commission. LAMOST is operated and managed by the National Astronomical Observatories, Chinese Academy of Sciences. 

Funding for the Sloan Digital Sky Survey IV has been provided by the Alfred P. Sloan Foundation, the U.S. Department of Energy Office of Science, and the Participating Institutions. SDSS-IV acknowledges
support and resources from the Center for High-Performance Computing at
the University of Utah. The SDSS web site is www.sdss.org.

SDSS-IV is managed by the Astrophysical Research Consortium for the 
Participating Institutions of the SDSS Collaboration including the 
Brazilian Participation Group, the Carnegie Institution for Science, 
Carnegie Mellon University, the Chilean Participation Group, the French Participation Group, Harvard-Smithsonian Center for Astrophysics, 
Instituto de Astrof\'isica de Canarias, The Johns Hopkins University, 
Kavli Institute for the Physics and Mathematics of the Universe (IPMU) / 
University of Tokyo, Lawrence Berkeley National Laboratory, 
Leibniz Institut f\"ur Astrophysik Potsdam (AIP),  
Max-Planck-Institut f\"ur Astronomie (MPIA Heidelberg), 
Max-Planck-Institut f\"ur Astrophysik (MPA Garching), 
Max-Planck-Institut f\"ur Extraterrestrische Physik (MPE), 
National Astronomical Observatories of China, New Mexico State University, 
New York University, University of Notre Dame, 
Observat\'ario Nacional / MCTI, The Ohio State University, 
Pennsylvania State University, Shanghai Astronomical Observatory, 
United Kingdom Participation Group,
Universidad Nacional Aut\'onoma de M\'exico, University of Arizona, 
University of Colorado Boulder, University of Oxford, University of Portsmouth, 
University of Utah, University of Virginia, University of Washington, University of Wisconsin, 
Vanderbilt University, and Yale University.
\end{acknowledgements}
\bibliographystyle{aa} 
\bibliography{ref.bib} 

\end{document}